\documentclass[a4paper,fleqn,usenatbib]{mnras}

\usepackage{mathptmx}

\usepackage[T1]{fontenc}
\usepackage{ae,aecompl}

\usepackage{graphicx}	
\usepackage{amsmath}	
\usepackage{amssymb}	
\usepackage{amsfonts}
\usepackage{pdflscape}
\usepackage{rotating}
\usepackage{natbib}
\usepackage{txfonts}
\usepackage{longtable}
\usepackage[small,bf]{caption}
\usepackage{multirow}
\usepackage{xspace}

\usepackage{epstopdf}

\newcommand\simlt{\lower.5ex\hbox{$\; \buildrel < \over \sim \;$}}

\newcommand{\gray}{$\gamma$-ray\xspace}
\newcommand{\grays}{$\gamma$-rays\xspace}

\title[Modeling HE flares of PKS 1441+25]{High Energy Gamma-Ray Emission From PKS 1441+25}
\author[N.~Sahakyan and S.~Gasparyan]{
N. Sahakyan$^{1,2}$ \thanks{E-mail: narek@icra.it}
and
S. Gasparyan$^{1}$\\
$^{1}$ ICRANet-Armenia, Marshall Baghramian Avenue 24a, Yerevan 0019, Republic of Armenia\\
$^{2}$ ICRANet, Piazza della Repubblica 10, I-65122 Pescara, Italy
}

\date{Accepted XXX. Received YYY; in original form ZZZ}

\pubyear{2017}

\begin{document}
\label{firstpage}
\pagerange{\pageref{firstpage}--\pageref{lastpage}}
\maketitle

\begin{abstract}
We present the $\gamma$-ray observations of the flat-spectrum radio quasar PKS 1441+25 (z=0.939), using the {\it Fermi} large Area Telescope data accumulated during January - December 2015. A $\gamma$-ray flare was observed in January 24, when the flux increased up to $(2.22\pm0.38)\times10^{-6}\;{\rm photon\:cm^{-2}\:s^{-1}}$ with the flux-doubling time scale being as short as $\sim1.44$ days. The spectral analysis shows that from April 13 to April 28, 2015 the MeV-to-GeV photon index has hardened and changes in the range of $\Gamma=(1.73-1.79)$ for most of the time. The hardest photon index of $\Gamma=1.54\pm0.16$ has been observed on MJD 57131.46 with $11.8\sigma$ which is not common for flat-spectrum radio quasars. For the same period the \gray spectrum shows a possible deviation from a simple power-law shape, indicating a spectral cutoff at $E_{\rm cut}=17.7\pm8.9$ GeV. The spectral energy distributions during quiescent and flaring states are modeled using one-zone leptonic models that include the synchrotron, synchrotron self Compton and external inverse Compton processes; the model parameters are estimated using the Markov Chain Monte Carlo method. The emission in the flaring states can be modeled assuming that either the bulk Lorentz factor or the magnetic field has increased. The modeling shows that there is a hint of hardening of the low-energy index ($\sim1.98$) of the underlying non-thermal distribution of electrons responsible for the emission in April 2015. Such hardening agrees with the $\gamma$-ray data, which pointed out a significant $\gamma$-ray photon index hardening on April 13 to 28, 2015.
\end{abstract}




\section{INTRODUCTION}\label{sec1}
Recent observations in the \gray band ($\geq 100$ MeV) show that the extragalactic \gray sky is dominated by emission from blazars - an extreme class of Active Galactic Nuclei (AGNs) which have jets that are forming a small angle with respect to the line of sight \citep{urry}. Blazars are known to emit electromagnetic radiation in almost all frequencies that are currently being observed, ranging from radio to Very High Energy (VHE; $>$ 100 GeV) \gray bands. The broadband spectrum is mainly dominated by non-thermal emissions produced in a relativistic jet pointing toward the observer. Due to small inclination angle and large bulk motion, the emission from blazars is affected by relativistic beaming which has enormous effects on the observed luminosities. Indeed, the observed luminosity ($L_{\rm obs}$) is related to the emitted luminosity ($L_{\rm em}^\prime$) as $L_{\rm obs}=\delta^{3+\alpha}\:L_{\rm em}^\prime$. If so, the observed luminosity can be thus amplified by a factor of thousands or even more (usually $\delta\geq10$). Such amplification makes it possible to detect emission even from very distant blazars.\\
A key feature of the non-thermal emission from blazars is the distinct variability at all frequencies (with different variability time scales - from years down to a few minutes). The shortest variability time scales are usually observed for the highest energy band; an example is the minute scale variability of PKS 2155-304 \citep{aharpks} and IC 310 \citep{alek310} which implies that the emission is produced in a very compact region. Therefore, by observing blazars one gets a unique chance to investigate the jet structure on sub-parsec scales.\\
By their emission line features blazars are commonly grouped as BL Lacertae objects (BL-Lacs) and Flat-Spectrum Radio Quasars (FSRQs) \citep{urry}. BL Lacs have weak or no emission lines, while FSRQs have stronger emission lines. The difference in the emission-line properties of FSRQs and BL Lacs may be connected with that in the properties of accretion in these objects \citep{ghisellini2009}.\\
The multiwavelength observations of blazars  have shown that their Spectral Energy Distribution (SED) has two broad non-thermal peaks - one at the IR/optical/UV/X-ray and the other at the Higher-Energy (HE; $>$ 100 MeV) \gray band. The low-energy peak is believed to be due to the non-thermal synchrotron emission of relativistic electrons while the origin of the second component is still debated. One of the most widely accepted theories for the second peak is that it is produced from Inverse Compton (IC) scattering of low energy synchrotron photons (Synchrotron Self Compton; SSC) \citep{ghisellini, maraschi, bloom} which often successfully explains the emission from BL-Lacs \citep{finke}. Besides, the photons from the regions outside the jet may serve as seed photons for IC scattering - External Compton (EC) models which are used to model the emission from FSRQs. The external photon field can be dominated either by the photons reflected by BLR \citep{sikora} or by photons from a dusty torus \citep{blazejowski,ghiselini09}. Domination of one of the components mostly depends on the localization of the emitting region; for example, if the energy dissipation occurs within BLR then the observed HE emission is mostly due to IC scattering of BLR reflected photons, otherwise, if the emitting region is far from the central source, then the IC scattering of torus photons will dominate. SSC and EC models assume that the emission is produced by the same population of electrons, though up to now it is not clear whether  it is produced in the same part of the jet or by different electron populations. Alternatively, the HE emission can be explained by the interaction of energetic protons; e.g., a significant fraction of the jet power goes for acceleration of protons so that they reach the threshold for pion production \citep{mucke1,mucke2}.\\
The majority of the blazars detected in VHE \gray band are high-frequency-peaked BL Lacs for which the synchrotron bump is in the UV/X-ray bands.  In addition to BL Lacs, there are also 5 FSRQs detected in the VHE \gray band which is rather surprising, since the Broad Line Region (BLR) structure of these objects, which is rich in optical-UV photons, makes these environments strongly opaque to VHE \grays \citep{liu, poutanen}. Moreover, FSRQs have a relatively steep photon index in the energy range of $>100$ MeV as was observed with the Fermi Large Area Telescope ({\it Fermi} LAT) which does not make them as strong emitters of VHE \gray photons. Detection of FSRQs in the VHE \gray band is challenging for the near-black-hole dissipation scenarios;  it assumes that the \grays are most likely produced farther from the central source, outside the BLR, where the dominant photon field is the IR emission from the dusty torus. Typically, the temperature of torus photons $\sim 10^3$ K is lower than that of the photons reflected in the BLR $\sim 10^5$  K, and, in principle, VHE photons with energy up to $\sim$ 1 TeV can escape from the region. Thus, the observations of FSRQs in VHE \gray band provide an alternative view of blazar emission as compared to BL Lacs. Moreover, since FSRQs are more luminous than BL Lacs, they could, in principle, be observed at greater distances. Indeed, the farthest sources detected in the VHE \gray band are the FSRQs at a redshift of $z\geq0.9$ (e.g., PKS 1441+25 \citep{abeysekara, ahnen} and S3 0218+35 \citep{ahnen16}). That is why FSRQs are ideal for estimation of the intensity of Extragalactic Background Light (EBL) through the absorption of VHE photons when they interact with the EBL photons \citep{coppi, made}.\\
Among FSRQs, PKS 1441+25 is one of the most distant sources detected so far at z=0.939 \citep{shaw}. In April 2015 both VERITAS and MAGIC collaborations announced the detection of VHE \grays from PKS 1441+25 (with up to 250 GeV photons) \citep{mirzoyan,mukherjee}. A strong emission from the source had been detected on April 20 to 27, 2015. During the same period, the source had been also observed with the telescopes Swift and NuSTAR. The origin of the multiwavelength emission from PKS 1441+25 observed in April is modeled assuming the emission region is beyond the BLR, and the emission in the VHE \gray band is mostly due to the IC scattering of the dusty torus photons \citep{abeysekara,ahnen}. Moreover, the large distance to PKS 1441+25 allowed to indirectly probe the EBL absorption at redshifts up to $z\sim1$ with  the help of ground-based \gray instruments.\\
In the theoretical interpretation of the multiwavelength emission from blazars, the size/location of the emitting region, magnetic field and electron energy distribution are uncertain. Only during flaring periods some of the unknown parameters can be constrained based on the observations in different bands. The observations of PKS 1441+25 during the bright period in April 2015 by different instruments provide us with data on the maximums of the emitting components (Swift UVOT/ ASAS-SN and {\it Fermi} LAT) as well as on the transition region between these components in the energy range from 0.3 to 30 keV (Swift XRT and NuSTAR) \citep{abeysekara}. Similar data (up to HE \gray band) are available also from the observations carried out on January 06 to 28, 2015, which is the period of the large flare that was observed with {\it Fermi} LAT. Thus, by modeling the emission in these two periods and estimating the parameter space that describes the underlying particle distribution responsible for the emission through the Markov Chain Monte Carlo (MCMC) technique, one can investigate and explore particle acceleration/emission processes and jet properties in these two significant flaring periods which are crucial for understanding the origin of the flares. This motivated us to have a new look at the origin of the multiwavelength emission from PKS 1441+25, using currently available data from Swift, NuSTAR and {\it Fermi} LAT.\\
This paper is structured as follows. The results of the spectral and temporal analysis of the {\it Fermi} LAT data are presented in Section \ref{sec2}. The broadband SED modeling with MCMC technique is presented in Section \ref{sec3} and discussion and conclusions are presented in Section \ref{sec4}.
\section{{\it Fermi} LAT DATA ANALYSIS}\label{sec2}
The large Area Telescope on board the {\it Fermi} satellite is a pair-conversion telescope sensitive to \grays in the energy range from 20 MeV to 300 GeV. It constantly scans the whole sky every 3 hours already more than 8 years. More details about {\it Fermi} LAT can be found in \citet{atwood09}.\\
In the present paper, for spectral analysis we use the publicly available data acquired in the periods from January 06 to 28 and from April 15 to 26, 2015. These two periods have been picked, because they are contemporaneous with the Swift XRT observations of the source \citep{abeysekara}. The data were analyzed with the standard {\it Fermi} Science Tools v10r0p5 software package released on May 18, 2015 available from the {\it Fermi} Science Support Center \footnote{http://fermi.gsfc.nasa.gov/ssc/data/analysis/software/}. The latest reprocessed PASS 8 events and spacecraft data are used with the instrument response function P8R2\_ SOURCE\_ V6 . We have downloaded photons in the energy range from 100 MeV to 100 GeV from a region of interest defined as a circle of a $20^{\circ}$ radius centered at the \gray position of PKS 1441+25 (RA, Dec) = (220.996, 25.039) \citep{acero15}. Only the events with higher probability of being photons ({\it evclass=128 evtype=3}) have been considered in the analysis. A cut on the zenith angle of $90^{\circ}$ is applied to reduce contamination from the Earth-limb \grays produced by cosmic rays at their interaction with the upper atmosphere. The model file, describing the region of interest, contains point sources from the {\it Fermi} LAT third source catalog \citep{acero15} (3FGL) within $25^{\circ}$ from the target, as well as contains Galactic {\it gll\_ iem \_ v05\_ rev1} and isotropic {\it iso\_source\_v05} diffuse components. All point-source spectra were modeled with those given in the catalog, allowing the photon index and normalization of the sources within $20^{\circ}$ to be free in the analysis. Also, the normalization of diffuse background components was not fixed.
\begin{figure}
   \includegraphics[width=0.5 \textwidth]{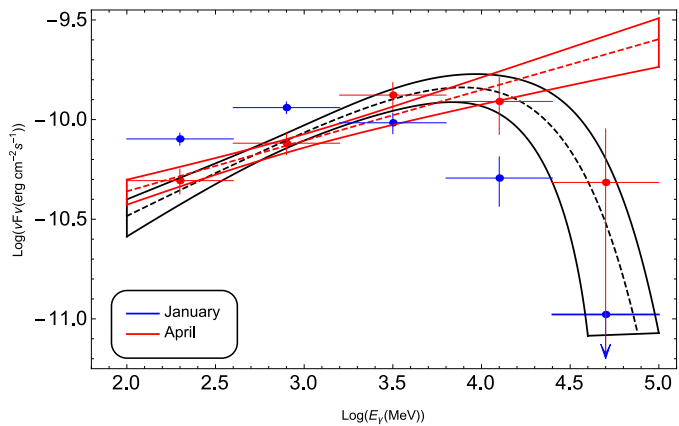}
   \caption{The \gray spectrum of PKS 1441+25 above 100 MeV averaged over the {\it Fermi} LAT observations in January (blue) and April (red).}
    \label{fg1}
\end{figure}
\subsection{Spectral analysis}\label{sec2.2}
In order to find the best matches between spectral models and events, an unbinned likelihood analysis is performed with {\it gtlike}. The PKS 1441+25 spectrum has been initially modeled as a power-law function where the normalization and the power-law index are taken as free parameters. The best fit parameters obtained with {\it gtlike} analysis are presented in Table \ref{table:param} and the corresponding spectrum is shown in Fig. \ref{fg1} (blue and red data for January and April, respectively). The spectrum is calculated by separately running {\it gtlike} for 5 energy bands equal on a log scale.\\
\begin{table}
 \centering
       \begin{tabular}{lcc}
         \hline
         \hline
           Parameter Name & Blue & Red  \\ \hline
          Flux (${\rm photon\:cm^{-2}\:s^{-1}}$)    & $(5.89\pm0.30)\times10^{-7}$ & $(3.63\pm0.36)\times10^{-7}$ \\
         $\alpha$    & 1.99$\pm$0.04 & 1.74$\pm$0.06\\
         TS    & 2174 & 910 \\
          \hline
         \end{tabular}
        \caption{The best parameters obtained with {\it gtlike} for power-law modeling. For each time period, photon flux in the range $0.1-100$ GeV, photon index and detection significance are presented.}
         \label{table:param}
\end{table}
\begin{figure*}
  \centering
    \includegraphics[width=1\textwidth]{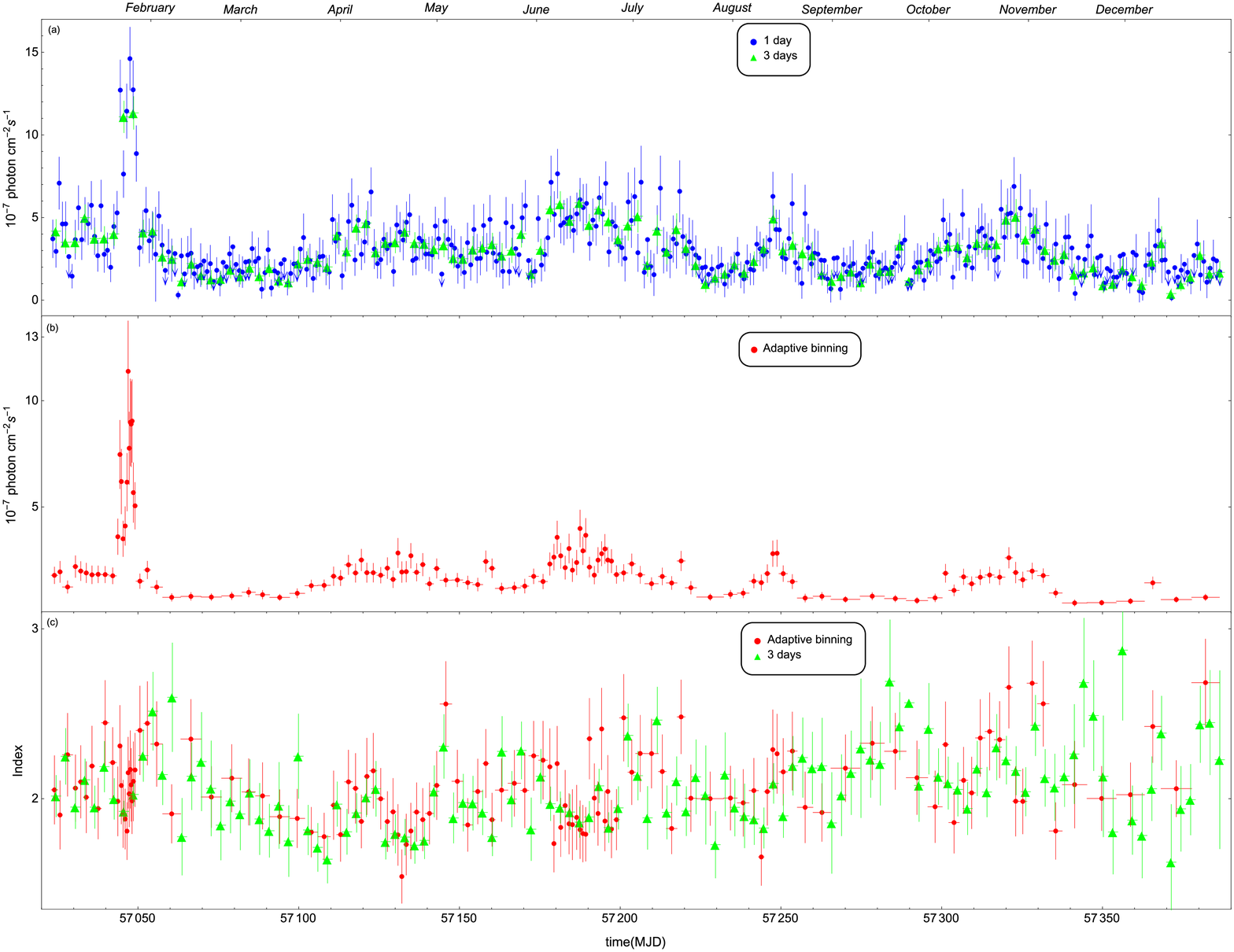}
    \caption{ The \gray light curve of PKS 1441+25 from January to December 2015 (a). The bin intervals correspond to 1- day (blue data) and 3-days (green data). The light curve obtained by adaptive binning method assuming 20 \%  of uncertainty  is presented in red (b). The change of photon index for 3-day binning (green) and with adaptive binning method are shown in (c).}%
    \label{fg2}
\end{figure*}
\begin{figure}
  \centering
    \includegraphics[width=0.45\textwidth]{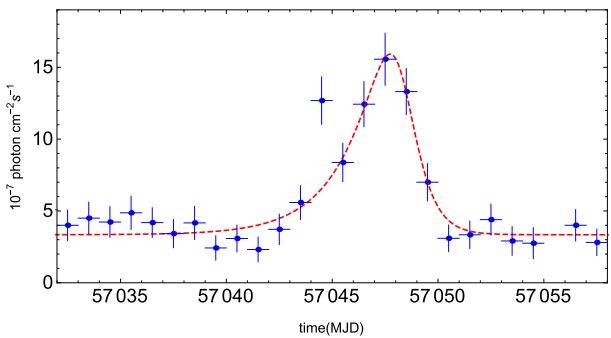}
    \caption{The light curve's sub interval that covers a major flaring period. The red dashed line shows the flare fit with Eq. \ref{func}.}%
    \label{fg2_fit}
\end{figure}
The fluxes presented in Table \ref{table:param} significantly exceed the averaged flux given in 3FGL ($\approx1.28\times10^{-8}\:{\rm photon\:cm^{-2}\:s^{-1}}$) \citep{acero15}. The photon index estimated in January 2015 is consistent with the value reported in 3FGL $\alpha=2.13$ (averaged over 4 years of observations); however, a relative hardening of $\alpha=1.74\pm0.06$ is observed in April, which is rarely observed for FSRQs. Moreover, we note an indication of deviation of the power-law model with respect to the data above 10s of GeV energies observed in April (red bowtie plot in Fig. \ref{fg1}). In order to check for a statistically significant curvature in the spectrum, an alternative fit of the power-law with an exponential cutoff function in the form of $dN/dE\sim E_{\gamma}^{-\alpha}\:\times Exp(-E_{\gamma}/E_{cut})$ is done, which results in $\alpha=1.56\pm0.1$ and $E_{\rm cut}=17.7\pm8.9$ GeV (black bowtie plot in Fig. \ref{fg1}). The power-law and cutoff models are compared with a log likelihood ratio test: the TS is twice the difference in the log likelihoods, which gives 8 for this case. Note that the TS probability distribution can be approximated by a $\chi^2$ distribution with 1 degree of freedom (dof) corresponding to the difference of the dof between the two functions. The results give $P(\chi^2)=0.0046$, which again indicates a deviation from a simple power-law function. The best-fit cutoff power-law function is shown as a black bowtie line in Fig. \ref{fg1}. However, $2.8\:\sigma$ is not a high enough significance to claim for a statistically significant curvature although it is as high as $3.86\:\sigma$ if the data collected during the whole month of April are considered.
\subsection{Temporal analysis}\label{sec2.3}
In order to investigate the size of the \gray emitting region, light curves with different time binning are generated. A characteristic timescale for flux variation $\tau$ would limit the (intrinsic) size of the emission region to $R\leq c\times \delta \times \tau/ (z+1)$. Thus, it is crucial to do a variability analysis in order to distinguish between different emission processes.\\
The light curve of PKS 1441+25 for the period from January to December 2015 has been calculated by the {\it gtlike} tool, applying the unbinned likelihood analysis method. $(0.1-100)$ GeV photons from a region with a $10^{\circ}$ radius centered on the position of PKS 1441+25 are used in the analysis with the appropriate quality cuts applied as in the previous case. During the analysis, in order to reduce the uncertainty in the flux estimations, in the model file the photon indices of all background sources are fixed to the best guess values. Two different sets of light curves are calculated, considering the power-law index of PKS 1441+25 as being fixed and then as free. Since no variability is expected for the background diffuse emission, the normalization of both background components is also fixed to the values obtained for the whole time period.\\
The \gray light curve of PKS 1441+25 obtained with one-day and three-days binning is presented in Fig. \ref{fg2} (a) (blue and green data respectively). In the light curve there can be identified several periods when the flux was in high as well as in quiescent states. A major increase of the \gray flux had been detected in the period from January 21 to 28, 2015, with a daily averaged maximum of $(1.55\pm0.18)\times10^{-6}\;{\rm photon\:cm^{-2}\:s^{-1}}$ observed on January 25, 2015. Unfortunately the peak flare of January 25 was not observed by Swift. The \gray photon index evolution in time in a three-day long binning is shown in Fig. \ref{fg2} (c) with green data (three-day long binning is used since the photon index uncertainties are less than in one-day binning). During the flaring period the photon index is $1.9-2.0$. Also an increase in the flux can be noticed around January 22nd which lasted just one day. In order to check if this brightening is statistically significant, light curves with denser time sampling (half a day and 4 hours) are generated. However, the corresponding flux increase is within the uncertainty of the surrounding bins, while the peak of the flux around 25th of January is present in both light curves. In addition, a substantial increase in the \gray flux was observed in April, from June to about July 15, mid August and around October-November; but the maximum flux intensity was lower as compared with that observed during the strong \gray outburst of January 21 to 28 (Fig. \ref{fg2}). The active state in April is the period when PKS 1441+25 was observed by MAGIC on MJD 57130-57139 and VERITAS on MJD 57133-57140 \citep{abeysekara,ahnen}. The \gray light curve with three-day binning shows that, between MJD 57125.56- 57140.64 (from April 13 to April 28, 2015), the \gray photon index is significantly harder, $\Gamma=(1.73-1.79)$. It implies that during the observations in the VHE band the source was in a state characterized by a hard \gray photon index in the MeV-to-GeV range.\\ 
Next, in order to investigate the flux changes in time, and in particular in the flaring periods, the light curves have been generated by an adaptive binning method. In this method, the time bin widths are flexible and chosen to produce bins with constant flux uncertainty \citep{lott}. This method allows detailed investigation of the flaring periods, since at times of a high source flux, the time bins are narrower than during lower flux levels, therefore the rapid changes of the flux can be found. In order to reach the necessary relative flux uncertainty, the integral fluxes are computed above the optimal energies \citep{lott} which correspond to $E_0$ = 215.4 MeV in this case. \\
Adaptively binned light curves in the 215 MeV-300 GeV energy range with 20\% and 15\% uncertainties have been generated. Flare is present in both light curves. The light curve with 20\% flux uncertainty at each bin is presented in Fig. \ref{fg2} (b) with red color. It confirms all the features visible in the constant-bin-width light curve, but also allows us to investigate fast variability during high-flux states in greater detail. The first flare episode occurred during MJD 57043.30-57049.38, when the time width was less than $\sim15$ hours. A strong flaring period is observed around the 24th-25th of January. The flux peak of $(1.14\pm0.24)\times10^{-6}\;{\rm photon\:cm^{-2}\:s^{-1}}$ was observed on the 24th of January at 22:35 $pm$ in a bin with a half-width of 3.1 hours. The analysis of the data acquired in the mentioned period on the energies of $>100$ MeV results in a flux of $(2.22\pm0.38)\times10^{-6}\;{\rm photon\:cm^{-2}\:s^{-1}}$, which is the highest photon flux detected from this source. The data analysis for the entire flaring period (January 21-28) resulted in a flux of $(1.05\pm0.06)\times10^{-6}\;{\rm photon\:cm^{-2}\:s^{-1}}$ and a photon index of $\sim1.98\pm0.04$ \citep{sahakpks}. After MJD 57049.38, PKS 1441+25 was in its quiescent state, and the data should be accumulated for more than a day to reach 20 \% uncertainty. Then from  MJD 57109.89 to MJD 57143.91, PKS 1441+25 was again in its active state which was characterized by emission with a significantly hardened \gray photon index. Starting from MJD 57126.70 to MJD 57141.93, the photon index of PKS 1441+25 hardened and reached $\leq1.9$ most of the time. Measured within a few hours, the photon index kept varying from $\Gamma=1.73$ to $\Gamma=1.91$. The hardest photon index of $\Gamma=1.54\pm0.16$ was observed on MJD 57131.46 with $11.8\sigma$ and the data being accumulated for $\approx29$ hours. Other periods, when PKS 1441+25 was bright enough to be detected on sub-day scales, are MJD 57177.38- 57199.76 and MJD 57243.02-57249.39.  For the rest of the time the source was in its quiescent state and the data should be accumulated for a few days or even longer in order to detect the source. The analysis of the light curve with the new adaptive binning method for the first time allowed us to investigate the flaring activity of PKS 1441+25 with a sub-day resolution and to perform detailed investigation of the flux and photon index changes.\\
Furthermore, to derive the flare doubling timescales and understand the nature of the January flare, the light curve is fitted with an exponential function in the form of \citep{fit}
\begin{equation}
F(t)= F_{\rm c}+F_0\times\left(e^{\frac{t-t_{\rm 0}}{t_{\rm r}}}+e^{\frac{t_{\rm 0}-t}{t_{\rm d}}}\right)^{-1}
\label{func}
\end{equation}
where $t_{0}$ is the time of the maximum intensity of the flare ($F_0$), $F_c$ is the constant level present in the flare, $t_{\rm r}$ and $t_{\rm d}$ are the rise and decay time constants, respectively. The fit shows that the flare is best explained when $t_{0}=57048.25\pm0.18$, $t_r= 1.92 \pm 0.3$, $t_d= 0.72 \pm 0.1$ and $F_0=(22.6 \pm 1.4)\times 10^{-7}\mathrm{photon\:cm^{-2}s^{-1}}$. The fit of the flaring period is shown in Fig. \ref{fg2_fit} with a dashed red line. Using this technique, it is also possible to estimate the shortest time variability (flux doubling) defined by $\tau=2 t_{\rm r,d}$, corresponding to $\tau=1.44$ days which is used to put an important constraint on the radiative region size. We note that the previous PKS 1441+25 \gray emission studies with the {\it Fermi} LAT data that covered only the period in April did not allow to properly estimate the \gray emitting region size, while here the analysis of the flaring period in January allowed to constrain the flare doubling time which is necessary for constraining the \gray emission region size.
\section{BROADBAND SED MODELLING}\label{sec3}
It is hard to make theoretical modeling of the observed broadband SED because the structure of the central region of blazars is complex and the exact localization of emitting regions is unknown. The observed fast variability indicates compactness of the emitting region but its localization remains an open problem. Along the jet, the emission can be produced in different zones and depending on the distance from the central black hole different components can contribute to the observed emission \citep{sikora09}.
\subsection{Broadband SED}
The broadband SEDs of PKS 1441+25 for different periods are shown in Fig. \ref{sed} where with red and blue colors are the SED observed in January and April respectively, while the archival data from ASI science data center \footnote{http://tools.asdc.asi.it/SED/index.jsp} are shown with gray color. We note that during the high states, the second emission peak increased by intensity and shifted to HEs. This kind of change has already been observed during the flaring state of 4C+49.22 \citep{cutini} and PKS 1510-089 \citep{abdopks1510}.  During the flaring periods the low-energy component's  intensity increased as compared with the quiescent state; the increase in April exceeded that one observed in January (although the power-law photon index in the X-ray band ($\approx 2.3$) had been relatively constant during both observations). More evident and drastic is the change of the peak intensity of the low energy component; from January to April it increased by nearly an order of magnitude and as compared with the quiescent state it increased $\geq 15$ times. On the contrary, the peak of the second component (in the HE \gray band) is relatively constant, only the photon index in the MeV-GeV energy range is harder during the observations in April. The Compton dominance of the source is stronger and evident during the flaring periods which suggests that the density of the external photon fields significantly exceeds the synchrotron photon density ($U_{\rm ext}/U_{\rm syn}>>1$). \\
Such a strong amplification of the emission from blazars can be explained by means of introducing changes in the emission region parameters, e.g., in the magnetic field, emitting region size, bulk Lorentz factor and others, and/or particle energy distribution. In principle, all the parameters describing the emitting region can be changed at the same time if the flares are due to a global change in the physical processes in the jet, which also affect the jet dynamics and properties. However, usually, the change in one or two parameters is enough to explain the flares. An interesting study of the flaring activity in FSRQs as a result of changes in different parameters has been investigated in \citet{paggi}. Namely, the emission spectra evolution as a function of changes in different parameters (e.g., bulk Lorentz factor, magnetic field, accretion rate, etc.) is investigated. In the case of PKS 1441+25, during its flaring periods, both the low energy and HE components increased several times. The increase of the second component is most likely due to moving of the emitting region outside its BLR. In principle, there are two possibilities: {\it i)} either the emitting region moves faster due to increasing bulk Lorentz factor and leaves the BLR or ii) the bulk Lorentz factor is unchanged and only the emitting region is moving beyond the BLR. In the first case, since the external photon density in the comoving frame of the jet depends on the Doppler boosting factor, a strong increase in the Compton dominance will be observed. We note that the change of the bulk Lorentz factor will also affect the low energy component. In the second case, the flaring activity is due to the change of the location of the emitting region and due to the magnetic field amplification. Additional increase of the magnetic field from January to April is also evident when the low energy component kept increasing (this corresponds to the case shown in Fig. 1 (b) in \citep{paggi}). Accordingly, we discuss two possibilities. First, we assume that $\delta$ has increased from 10 in the quiescent to 18 in the flaring periods, and then we assume that it was constant ($\delta=18$) in both periods. These values are below and above the estimated mean bulk Lorentz factor of FSRQs obtained from the analysis of a large sample of \gray emitting FSRQs \citep{ghistav}. The emission region size can be estimated through the observed variability time scale $\tau=1.44$ d implying that $R_b\leq\delta\:c\:\tau/(1+z)\approx3.5\times10^{16}$ cm when $\delta=18$ and $R_b=1.92\times10^{16}$ cm when $\delta=10$.
\subsection{Theoretical modeling}
We attempt to fit the SEDs in the high states of January and April as well as in the quiescent state. Even if a quiescent state SED is constrained with non-simultaneous data, its modeling provides an insight into the dominant physical processes which are constantly present in the jet but are covered by the flaring components during the high states. 
We modeled the PKS 1441+25 SED for high and quiescent states in the framework of single-zone leptonic models that include the synchrotron, SSC, and EC processes. The emission region (the "blob"), assumed to be a sphere with a radius of $R$ which is moving with a bulk Lorentz factor of $\Gamma$, carries a magnetic field with an intensity of $B$ and a population of relativistic electrons. The blob velocity makes a small angle with respect to the line of sight, so the emission is amplified by a relativistic Doppler factor of $\delta$. The energy spectrum of the population of electrons in the jet frame, which is responsible for the non-thermal emission is assumed to have a broken power-law shape:
\begin{equation}
      N_e^{\prime}(E_{e}^\prime)=
      \begin{cases}
	N_0^\prime \left(\frac{E_{e}^\prime}{m_ec^2}\right)^{-\alpha_1} & E_\mathrm{e,min}^\prime \leqslant E_{e}^\prime \leqslant E_\mathrm{br}^\prime \\
	N_0^\prime \left(\frac{E_{br}^\prime}{m_ec^2}\right)^{\alpha_2-\alpha_1} \left(\frac{E_{e}^\prime}{m_ec^2}\right)^{-\alpha_2} & E_\mathrm{br}^\prime \leqslant E_{e}^\prime \leqslant E_{e,max}^\prime
      \end{cases}
      \label{BPL}
      \end{equation}
      \noindent
where $N_{0}^\prime$ is connected with the total electron energy $U_{\rm e}=\int_{\rm E_{min}^\prime}^{\rm E_{\rm max}^\prime}E_{e}^\prime N_e(E_{e}^\prime)dE_e^\prime$, $\alpha_{1}$ and $\alpha_{2}$ are the low and high indexes of electrons correspondingly below and above the break energy $E_{\rm br}^\prime$, and $E_{\rm min}^\prime$ and $E_{\rm max}^\prime$ are the minimum and maximum energies of electrons in the jet frame, respectively. The electron spectrum given in Eq. \ref{BPL} is naturally formed from the cooling of relativistic electrons \citep{kardashev,inoue}.\\
The low-energy (from radio to optical/X-ray) emission is due to the synchrotron emission of electrons with an energy spectrum as given by Eq. \ref{BPL} in a homogeneous and randomly oriented magnetic field. For the quiescent state we assume the energy dissipation occurs close to the central source region and it is explained as an IC scattering of synchrotron photons (SSC). Instead the high state emission is dominated by that from a region well outside the BLR in order to avoid the strong absorption of photons with energies $\geq\:100\:{\rm GeV}$ (similar assumptions have been already made in \citep{abeysekara,ahnen}). In this case the dominant external photon field is the IR radiation from the dusty torus which, as we assume, has a blackbody spectrum with a luminosity of $L_{\it IR}=\eta\:L_{\it disc}$ ($\eta=0.6$, \citep{ghisellini2009}) and a temperature of $T=10^{3}$ K and fills a volume that for simplicity is approximated as a spherical shell with a radius of $R_{IR}=3.54\times10^{18}\:(L_{\rm disc}/10^{45})^{0.5}$ cm \citep{nenkova}. The disc luminosity is estimated from the BLR luminosity, $L_{\rm disc}=10\times L_{\rm BLR}\approx 2\times10^{45}{\rm erg\:s^{-1}}$ \citep{xiong}.
\subsection{Fitting technique}
In order to constrain the model parameters more efficiently, we employed the MCMC method, which enables to derive the confidence intervals for each model parameter. For the current study we have modified the {\it naima} package \citep{zabalza} which derives the best-fit and uncertainty distributions of spectral model parameters through MCMC sampling of their likelihood distributions. The prior likelihood, our prior knowledge of the probability distribution of a given model parameter and the data likelihood functions are passed onto the emcee sampler function for an affine-invariant MCMC run. In addition, there are multiple simultaneous walkers which improve the efficiency of the sampling and reduce the number of computationally expensive likelihood calls. We run the sampling with 64 simultaneous walkers, for 100 steps of burn-in, and 100 steps of run. In the parameter sampling, the following expected ranges are considered: $1.5\leq(\alpha_{1,2})\leq10$, $0.511\:{\rm MeV}\leq E^\prime_{(br,\:min,\:max)}\leq1\:{\rm TeV}$, and $N_0$ and $B$ are defined as positive parameters. The synchrotron emission is calculated using the parameterization of the emissivity function of synchrotron radiation in random magnetic fields presented in \citet{aharsyn} while the IC emission is computed based on the monochromatic differential cross section of \citet{aharatoy}.
\begin{figure}
   \centering
  \includegraphics[width=0.5 \textwidth]{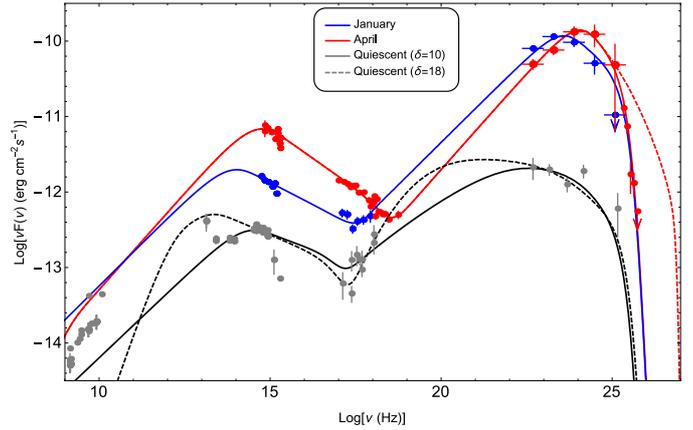}
   \caption{The broadband SED of PKS 1441+25 for January (red),  April (blue) and for the quiescent state (gray). The blue, red and gray lines are the models fitting the data with the electron spectrum given by Eq. \ref{BPL} for January, April and for the quiescent state, respectively. The model parameters are presented in Table \ref{table_fit}. The UV-X-ray and VHE \gray data observed in January and April are from \citet{abeysekara} and HE \gray data ({\it Fermi} LAT) are from this work.}
    \label{sed}
\end{figure}
\begin{table*}
\footnotesize
\renewcommand{\arraystretch}{1.7}
\begin{center}
\caption{Model parameters.}
\label{table_fit}
\begin{tabular}{lccccc}
\hline
& Parameter & Quiescent  & Quiescent & January & April \\
\hline
Doppler factor & $\delta$ & 10 & 18 & 18 & 18\\
Normalization of electron distribution & $N_0^\prime\times10^{48}\:{\rm eV^{-1}}$ & $10.68_{-2.64}^{+3.09}$ & $43.44_{-7.76}^{+6.59}$  & $23.83^{+8.11}_{-7.32}$   & $6.12^{+1.67}_{-1.56}$ \\
Low-energy electron spectral index & $\alpha_1$ & $2.14\pm0.04$ & $2.09_{-0.04}^{+0.03}$ & $2.10_{-0.05}^{+0.04}$   & $1.98\pm 0.03$ \\
High-energy electron spectral index & $\alpha_2$ & $3.39_{-0.14}^{+0.27}$ & $3.38\pm0.06$ &$3.46\pm0.06$ & $3.64\pm 0.01$  \\
Minimum electron energy & $E^\prime_{\rm min}$ (MeV) & $1.84_{-1.23}^{+1.75}$ & $286.37_{-25.39}^{+30.64}$& $1.97^{+0.31}_{-0.34}$  & $4.16^{+1.00}_{-1.86}$ \\
Break electron energy &$E^\prime_{\rm br}$ (GeV) & $2.83_{-0.31}^{+0.51} $ & $1.11_{-0.12}^{+0.14}$ & $1.62^{+0.23}_{-0.15}$ & $3.11^{+0.15}_{-0.23}$ \\
Maximum electron energy & $E^\prime_{\rm max}$ (GeV) & $46.27_{-13.76}^{+49.74}$ & $82.32_{-17.14}^{+13.47}$ & $127.82^{+26.74}_{-24.75}$  & $202.79^{+21.2}_{-14.6}$ \\
Magnetic field & B [G] & $0.19\pm0.013$ &  $0.046\pm0.002$ & $0.11^{+0.005}_{-0.004}$         & $0.18^{+0.009}_{-0.006}$\\
Jet power in magnetic field & $L_{B}\times10^{43}$ erg s$^{-1}$ & $0.49$ & 0.31  & $1.71$ & $4.51$ \\
Jet power in electrons & $L_{e}\times10^{45}$ erg s$^{-1}$ & 2.11 & 4.07 & $9.60$ & $4.47$\\
\end{tabular}
\end{center}
\end{table*}
\subsection{SED modeling and results}
The results of SED modeling are shown in Fig. \ref{sed} with the corresponding parameters in Table \ref{table_fit}. The radio emission is due to the low-energy electrons which are accumulated for longer periods, that is why, the radio data are treated as an upper limit for the purposes of our modeling. To have an indication of a change in the energetic contents of the jet, as well as of changes in the radiating particle distribution, first we try to fit the SED in a quiescent state which is modeled assuming two different Doppler boosting factors. The gray solid line in Fig. \ref{sed} shows the synchrotron/SSC emission assuming that the jet Doppler boosting factor is $\delta=10$, and the gray dashed line is the case of $\delta=18$. In case of $\delta=10$, as the emitting region size is as small as $R_b=1.92\times10^{16}$ cm, the magnetic field should be as strong as $B=0.19$ G to account for the observed data, while at $\delta=18$ the magnetic field is much weaker, $B=0.046$ G. Also, the underlying electron distribution for the case of $\delta=10$ is characterized by a slightly higher break ($2.83$ GeV versus $1.11$ GeV) in order to account for the observed emission.\\
The emission in flaring periods is modeled assuming that the HE emission is entirely due to the IC scattering of external photons (Fig. \ref{sed}). In all calculations the absorption due to the EBL was taken into account using a model from \citet{franceschini} since a strong absorption is evident at $\geq\:100$ GeV (red dashed line in Fig. \ref{sed}). In both periods the HE electron spectral index is within the range of $\alpha_2\sim (3.46-3.64)$ which is required to explain the UV-X-ray data with a photon index of $\approx2.3$. The lack of low-energy data makes the precise estimation of the low energy electron index harder. Only the Swift XRT/NuSTAR data from the observation of the transition region between low and high energy components allows to define the parameters $E_{\rm min}$ and $\alpha_1$. The low energy electron index is in a typical range expected from shock acceleration theories, $\alpha_1\approx2$.\\
As distinct from the quiescent state, in order to explain the flaring activities,  both, the electron distribution and the magnetic field should be varying. We note that the magnetic field required for modeling of flaring periods, ($B\geq$ 0.11 G), is weaker than that one estimated in the quiescent state in case of $\delta=10$ ($B\sim 0.19$ G). Since the synchrotron emission depends on the total number of emitting electrons $N_{\rm e}$, $\delta$ and magnetic field strength $B$, in case of smaller $\delta$ (and emitting region size) the magnetic field should be stronger. Instead, when $\delta$ is constantly equal to 18 in both states, the magnetic field should be nearly $\sim$ 2.4 and $\sim$ 3.9 times stronger in January and April, respectively, in order to explain the observed data. As the synchrotron photon density is proportional to $B^2$, the increase in the magnetic field strength resulted in the observed increase of the synchrotron flux by a factor of $5.7$ and $15.3$ (Fig. \ref{sed}). In the modeling of the SEDs observed in the flaring periods of January and April, the magnetic field should be changed in accordance with the increase in the low energy component.  Since the emission in the HE band is dominated by the IC scattering of external photons, this component remains stable during those periods (this corresponds to Fig. 1 (b) \citet{paggi}).\\
\begin{figure}
   \centering
  \includegraphics[width=0.5 \textwidth]{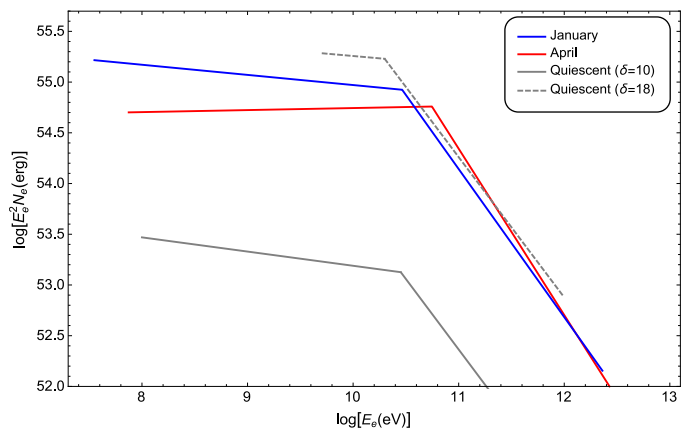}
   \caption{The electron spectra (broken power laws) obtained from the fit of the quiescent and flaring states of PKS 1441+25. Details on the parameter values are given in Table \ref{table_fit}.}
    \label{elect}
\end{figure}
The electron spectra obtained during the fit of SEDs in quiescent and flaring states are shown in Fig. \ref{elect}. It is clear the evolution of the electron spectra during the quiescent and high states. The low energy indexes of the underlying electron distribution in the quiescent state ($\delta=10$) are softer as compared with the flaring period (April). The total electron energy for modeling the emission in the quiescent period, when $\delta=18$, is almost of the same order as that in the flaring periods, which is expected, as the magnetic field is weaker, most of the jet energy is carried by particles. During the flaring periods, there are evident changes also in the underlying electron distribution.  The electron distribution best describing the data observed in April hints at {\it i)} hardening of the low energy index, {\it ii)} a higher break at $\sim 3.1$ GeV and maximum energies of $\sim203$ GeV. $E_{\rm br}$ and $E_{\rm max}$ are expected to shift, as the \gray spectrum observed in April is slightly inclined toward HEs, as compared with the January spectrum (see Fig. \ref{sed}). However, due to the large uncertainties in the estimations, especially for $\alpha_1$ (since the data in between $100$ keV and 100 MeV are missing), no definite conclusions can be drawn. For a statistically significant claim for hardening, there are required additional data in the energy range characterizing the rising part of the low and high energy components, which will allow to constrain $\alpha_1$ with higher confidence.  We note, however, that the significant hardening of the \gray emission observed in April (Fig. \ref{fg2} (c)) supports and strengthens the assumptions on the hardening of the low energy electron index.\\
Similar modeling of the SED of PKS 1441+25 observed in April has been already done in \citep{abeysekara,ahnen}, but it was done in a different manner. For example, in \citep{ahnen} the low electron energy index is fixed to be $\alpha_1=2$, a value expected from strong shock acceleration theories, while in our case all the parameters can vary in the fitting procedure. After having observed the hardening of the \gray photon index in April, we believe that exact estimation of $\alpha_1$ is important. Moreover, possible hardening or softening of $\alpha_1$ would point out the acceleration processes in the jet. However, the main difference in the modeling presented here, as compared with the previous ones, is the size of the emitting region (blob). They used larger blob size, $5\times10^{16}$ cm, in \citep{ahnen} and $4\times10^{17}$ cm in \citep{abeysekara}.  In our case, the modeling of the January flare time profile allowed us to constrain the emitting region size by $R_b\leq 3.5\times10^{16}\:(\delta/18)$ cm. Another difference with the previously reported parameters is that in our case the electron energy density is nearly 100 times higher than the magnetic field energy density. In \citep{abeysekara} $U_{e}/U_{B}=1.5$, which is related to the fact that much bigger emitting region size is used. We note that in \citep{ahnen}, where the considered blob size is similar to our case, they also found that $U_{e}/U_{B}\geq10$. Moreover, in our case the radius of the IR torus is derived from a different scaling law, which can cause additional difference in the estimation of the total energy. Despite using different approaches and parameters as compared with those used in the previous modelings, we note, that the main parameters for the underlying electron distribution obtained during April are similar to the previously reported values.
\section{DISCUSSION and CONCLUSIONS}\label{sec4}
In this paper, we present the results of the study of the \gray emission from PKS 1441+25 during January-December 2015. The data from the observations of a bright GeV flare in January allow us to estimate the emitting region size whereas the modeling of the broadband SED of PKS 1441+25 in January and April provided a chance to probe into the physical process during the flaring periods.\\
The \gray light curve generated with an adaptive binning method shows that the source entered its high activity state around MJD 57043.3 and the flux reached its maximum on January 24, when, within a few hours, the flux increases up to $F_{\gamma}(>\:100\:{\rm MeV})=(2.22\pm0.38)\times10^{-6}\;{\rm photon\:cm^{-2}\:s^{-1}}$. During this \gray { \bf brightening} the fit of the flare profile shows a slow rise and a fast decay trend with the shortest variability (flux doubling) time being $\tau_{\rm d}=1.44$ days. The rise of the flare can be attributed to the shock acceleration, whereas the decay phase cannot be explained by cooling of particles. Indeed, for the electrons that emit \grays with $\epsilon_\gamma=1$ GeV, as measured in the observer frame, the corresponding cooling timescale would then be $\sim(3\:m_e\:c/\sigma_T\:u_{\rm IR})\times(\epsilon_{\rm IR}(1+z)/\epsilon_\gamma)^{0.5}$ \citep{saito} which corresponds to 0.47 days in this case. This timescale is shorter than the observed e-folding decay timescales of the flares, implying that the observed flux decrease is related to the processes other than radiative losses.\\
After the flare on MJD 57049.38, the source is in its quiet state and the next increase in the flux is observed starting from MJD 57109.89.  Even if during this period, the flux amplitude is lower than one that observed in January, an interesting modification of  the \gray emission spectrum is observed. First, the \gray photon index hardened during MJD 57126.70-57141.93, it was $\leq1.9$. This period coincides with the one when VHE \grays from PKS 1441+25 were detected. The hardest \gray photon index, $\Gamma=1.54\pm0.16$, has been observed on MJD 57131.46 with a convincingly high detection significance of $11.8\sigma$. This photon index is unusual for FSRQs which are with an averaged photon index of $2.4$ in the third {\it Fermi} LAT AGN catalog (see Fig. 8 of \citet{ackermancatalog}). This photon index is even harder than the index of B3 1151+408 ($\Gamma=1.77$) which has the hardest photon indexes in the clean sample of {\it Fermi} LAT detected FSRQs. Although, hard photon indexes have been occasionally observed during rapid flaring events in FSRQs \citep{pacciani}. The observed hardening was perhaps related to the emission of new energetic particles that were either injected into the emitting region or re-accelerated. Next, the data analysis covering only the period in April shows that the \gray flux hints at a spectral curvature and a power-law with an exponential cut-off model is preferred over the simple power-law modeling assuming a break around $E_{\rm cut}=17.7\pm8.9$ GeV with a significance of $2.8\sigma$. Although the low statistics does not allow to claim for a statistically significant curvature in the spectrum, the \gray photon index observed in the VHE \gray band ($\sim5.4$, which corresponds to an intrinsic index of $3.4$ after correction for the EBL) strongly supports the presence of a break or a cut-off in the PKS 1441+25 spectrum around tens of GeV. Most likely, this break is defined by the break present in the radiating electron spectrum rather than is caused by the absorption within BLR \citep{poutanen} (otherwise the photons with $>$100 GeV would be strongly absorbed).
\subparagraph{The origin of multiwavelength emission:}
The SEDs observed during quiescent and flaring states are modeled using one-zone leptonic models and the model parameters are estimated using the MCMC method. The HE \gray emission observed in the flaring states can be explained by IC scattering of IR photons from the dusty torus whereas the SSC model gives a satisfactory representation of the data observed during the quiescent state.
The flares observed in January and April can be explained assuming there are changes in the bulk Lorentz factor or in the magnetic field. If the emitting region leaves the BLR region due to the increase of the bulk Lorentz factor (from $\delta=10$ to $\delta=18$), the Compton dominance will increase as it has been observed in the \gray band. Indeed in the flaring states, the IC to synchrotron luminosities ratio $L_{\gamma}/L_{\rm syn}\approx200$ and $\approx28$ in January and April, respectively as compared with that in the quiescent state $L_{\gamma}/L_{\rm syn}\approx(2-4)$. At the same time, the increase in the low energy component indicates that the magnetic field also increased between the flares in January and April \citep{paggi}. On the other hand, if the bulk Lorentz factor is unchanged ($\delta=18$), only the change in the emitting region location and amplification of the magnetic field can explain the multifrequency behavior observed during the flares. It is possible to  distinguish between these two scenarios, provided there are data in the hard X-ray or soft \gray band, as the modeling with $\delta=18$ predicts a higher flux in the hard X-ray band than when $\delta=10$ is assumed (gray dashed and solid lines in Fig. \ref{fg2}). Such data are missing in this case, making it hard to give exact interpretation of the origin of the flare. Anyway, physically reasonable parameters are used in both of these scenarios.\\ 
When comparing the electron parameters required for the modeling of the SEDs in January and April, we find a hint of possible hardening of the low energy electron index in April. We note, however, that no definite conclusions can be drawn since $\alpha_1$ is poorly constrained (due to missing or nonsufficient data). For all that, the April hardening of the \gray photon index in the MeV-GeV energy region supports our assumptions on hardening of the power-law index of the underlying electron distribution.
\subparagraph{Jet Energetics:} The jet power in the form of magnetic field and electron kinetic energy are calculated by $L_{B}=\pi c R_b^2 \Gamma^2 U_{B}$ and $L_{e}=\pi c R_b^2 \Gamma^2 U_{e}$, respectively, and are given in Table \ref{table_fit}. The jet power in the electrons changes in the range $(4.5-9.6)\times10^{45}\:{\rm erg\:s^{-1}}$ during the flares, while in the quiescent state it is of the order of $(2.1-4.1)\times10^{45}\:{\rm erg\:s^{-1}}$. Assuming one proton per relativistic electron (e.g., \citep{cellot,ghisbook}), the total kinetic energy in the jet is $L_{\rm kin}=8.02\times10^{47}\:{\rm erg\:s^{-1}}$ and $L_{\rm kin}=1.35\times10^{47}\:{\rm erg\:s^{-1}}$ for January and April, respectively.\\
The maximum \gray flux during the period of high activity is $(2.22\pm0.38)\times10^{-6}\;{\rm photon\:cm^{-2}\:s^{-1}}$ which corresponds to an isotropic \gray luminosity of $L_{\gamma}=1.22\times10^{49}\:{\rm erg\:s^{-1}}$ (using a distance of $d_L \approx 6112.8 $ Mpc). Likewise, the \gray luminosities in the periods of January and April were $L_{\gamma}=3.48\times10^{48}\:{\rm erg\:s^{-1}}$ and $L_{\gamma}=5.21\times10^{48}\:{\rm erg\:s^{-1}}$, respectively. Yet, at $\delta=18$ the total power emitted in the \gray band in the proper frame of the jet would be $L_{{\rm em,}\gamma}=L_{\gamma}/2\:\delta^2=1.89\times10^{46}\:{\rm erg\:s^{-1}}$ during the peak flux and would change within $L_{{\rm em,}\gamma}=(5.38-8.04)\times10^{45}\:{\rm erg\:s^{-1}}$ in January and April. These luminosity values account for only a small fraction ($\leq 6.7\%$) of the total kinetic energy of the jet. However, assuming that the standard radiative efficiency of the accretion disc $\eta_{\rm disc}\sim10\%$, the accretion power would be $L_{\rm acc}=2\times10^{46}\:{\rm erg\:s^{-1}}$. Thus during the flaring period the power emitted as \gray photons constitutes the bulk of the total accretion power $L_{{\rm em,}\gamma}/L_{\rm acc}\approx1$, while in January and April it made a substantial fraction of it - $L_{{\rm em,}\gamma}/L_{\rm acc}\approx(0.3-0.4)$; this is in a good agreement with the recent results by \citet{ghisellini14}, which showed that the radiative jet power in blazars is higher than (or of the order of) the accretion disk luminosity.\\
\\
The observations in both X-ray and \gray bands show that after the activity observed in January and April the emission from the source again enters a quiescent state.  A small increase in the \gray flux has been observed only in June, August and October-November 2015. Also, the UV/X-ray flux measured by Swift in May 2015 \citep{abeysekara} shows that the synchrotron component is weaker than it was in April. Thus, this indicates that the magnetic field in the emitting region started to decrease. In addition, in the \gray band, the flux slowly decreases down to a few times $10^{-7}\:{\rm photon\:cm^{-2}\:s^{-1}}$ for most of the time after August 2015, and the \gray photon index reaches its mean level. These point out that the emission from the blob outside the BLR region weakened, and the decrease of the Compton component shows that the emission responsible for the emission in the quiescent state (SSC) starts to dominate again. Since in this case the emission occurs close to the central source, due to the strong absorption, it is not expected to have emission of VHE \gray photons.\\
The multiwavelength observations of PKS 1441+25 during the flaring periods allowed us to investigate and discuss the changes that possibly took place in the jets and caused flaring activities. However, the parameters describing the underlying electron distribution below the break are poorly constrained, because the data describing the rising part of both low and HE components are missing. It did not allow us to exactly identify the processes responsible for the acceleration of particles in the jet. However, the future possible observations of flaring periods also in other energy bands will provide a chance to investigate the dominant particle acceleration processes.
\section*{acknowledgements}
This work was supported by the RA MES State Committee of Science, in the frames of the research project No 15T-1C375. Also, this work was made in part by a research grant from the Armenian National Science and Education Fund (ANSEF) based in New York, USA. We thank the anonymous referee for constructive comments that significantly improved the paper.
\bibliographystyle{mnras}
\bibliography{biblio} 
\label{lastpage}
\end{document}